\documentclass{ifacconf}

\usepackage{graphicx}      % include this line if your document contains figures
\usepackage{natbib}        % required for bibliography
\usepackage{amssymb,mathtools}
\usepackage{xcolor}
\usepackage{comment}
\usepackage{dirtytalk}
\graphicspath{{./Figures/}}
\begin{document}
	\begin{frontmatter}
		
		\title{Eco-Driving of Connected and Autonomous Vehicles with Sequence-to-Sequence Prediction of Target Vehicle Velocity} 

		\author[First]{Shobhit Gupta} 
		\author[Second]{Marcello Canova} 
		%\author[Third]{Third C. Author}
		
		\address[First]{Center for Automotive Research, The Ohio State University, Columbus, OH 43212 USA (e-mail: gupta.852@ osu.edu).}
		\address[Second]{Center for Automotive Research, The Ohio State University, Columbus, OH 43212 USA (e-mail: canova.1@ osu.edu).}
		
		\begin{abstract}                % Abstract of not more than 250 words.
			The Eco-Driving control problem seeks to perform fuel efficient speed planning for a Connected and Autonomous Vehicle (CAV) that can exploit information available from advanced mapping, and from Vehicle-to-Everything (V2X) communication. The ability of an Eco-Driving strategy to adapt in real time to variable traffic scenarios where surrounding vehicles can be either connected or unconnected is critical for further development and deployment of this technology in the transportation sector.
			
			In this work, the Eco-Driving strategy, formulated as a receding-horizon optimal control problem, is integrated with a target vehicle speed prediction model and solved via Dynamic Programming (DP) to determine the optimal speed trajectory in the presence of a human-driven target vehicle. An encoder-decoder architecture analyzes the patterns in the target vehicle velocity recorded over a historic window using a Gated-Recurrent-Unit (GRU) based encoder and generates an estimate of the future velocity trajectory using the GRU based decoder. A sensitivity study is done to analyze the effect of the historical and prediction windows on the accuracy of the velocity  predictor. The proposed Eco-Driving controller is evaluated through microscopic simulations using a traffic simulator.
		\end{abstract}
		
		
	\end{frontmatter}
	%===============================================================================
	
	\section{Introduction}
	\label{intro}
	The recent advancements in Connected and Automated Vehicles (CAVs) technologies have the potential to increase safety, driving comfort as well as fuel economy, by exploiting information on driving conditions ahead that become available via advanced navigation systems, Vehicle-to-Vehicle (V2V) and Vehicle-to-Infrastructure (V2I) communication (\cite{guanetti2018control}). Intuitively, the availability of such technologies could be exploited in a predictive controller to plan a speed trajectory that minimizes unnecessary acceleration and braking events, thereby improving driver comfort and fuel-efficiency (\cite{xu2018design}). Furthermore, information on future driving conditions could enhance energy efficiency gains in electrified powertrain, by optimally scheduling the use of the electric motors and onboard battery pack (\cite{guzzella2007vehicle}).
	
	In this scenario, Eco-Driving aims at exploiting information on future driving conditions to optimize speed planning and powertrain control for improving the fuel efficiency between an origin and destination. However, Eco-Driving algorithms pose challenges for the design of a predictive (\say{look-ahead}) controller, especially in presence of uncertain traffic environment (\cite{alam2014critical}). Another limitation of the existing literature in Eco-Driving is that the problem of speed trajectory optimization and powertrain control is often treated as two decentralized problems to reduce the computational complexity. In this case, the speed trajectory prediction of the target vehicle is incorporated in the speed planning problem, and the resulting ego vehicle speed trajectory is utilized as an external input for the sub-problem of powertrain control optimization (\cite{firoozi2018safe}).
	
	For real-world implementation of Eco-Driving controls, it is crucial to predict the presence of target vehicles as additional dynamic states, and incorporate them into the trajectory planning process. While some work has been done to incorporate a velocity prediction for a fully or partially connected target vehicle (see for instance \cite{sun2014velocity,kamal2015intersection}), however such assumptions might not be realistic in a partially connected environment where the ego vehicle might have access to only the current value of the target vehicle velocity.
	
	In addition, forecasting the speed trajectory of a human-driven vehicle (no automation) in urban driving is a very difficult task, as the vehicle's maneuvers may by affected by various dynamic factors, for instance traffic light Signal Phasing and Timing (SPaT), driver's experience, weather, etc. Earlier studies have assumed simple but unrealistic speed trajectories, such as constant speed and constant acceleration when no information is available (for example, \cite{lefevre2014comparison}). Auto-regressive models proved to be more realistic when compared to the constant acceleration case, but these prediction methods induce large errors in a prediction horizon, ultimately degrading the controller performance (\cite{hyeon2019influence}). More sophisticated model-based car-following models have been proposed by \cite{gupta2019enhanced,rajakumar2020benchmarking}, however these models require the look-ahead driving information of the human-driven target vehicle within a Line-of-Sight (LoS), which may not be available to the ego vehicle.  Data-driven methods such as Long-Short-Term-Memory (LSTM) have been recently studied for forecasting speed trajectories, since they have the ability of predicting highly non-linear time-series with minimal input information (\cite{altche2017lstm}). Although LSTM have proven to have high accuracy, they generally require large training time and are ideal for long-term prediction (\cite{altche2017lstm}).
	
\begin{figure}[t!]
		\centering
		\includegraphics[width=\columnwidth]{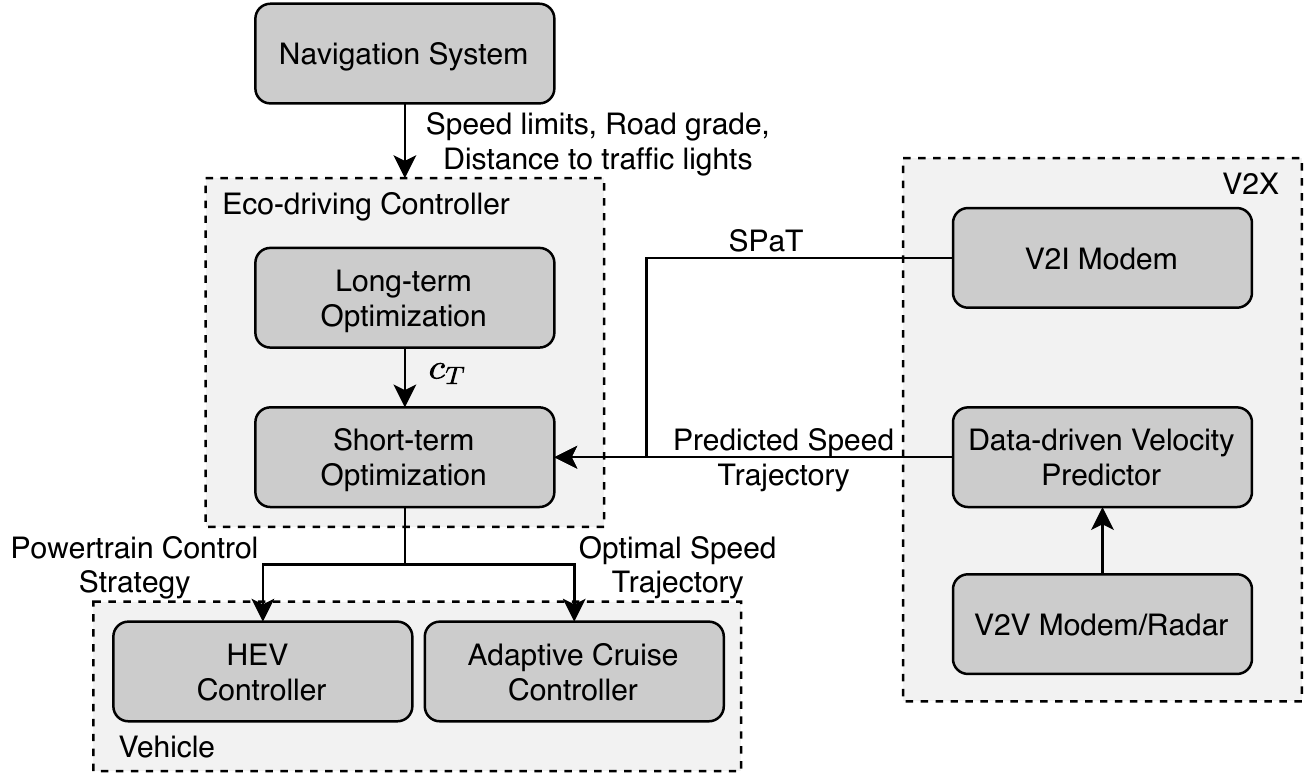}
		\caption{Control architecture of proposed Eco-Driving controller exploiting V2X technology in presence of partially connected, human-driven target vehicle.}
		\label{fig: eco-drving-architecture}
	\end{figure}

	This work proposes a unified Eco-Driving framework that performs co-optimization of vehicle velocity and powertrain control, as shown in Fig. \ref{fig: eco-drving-architecture}. The proposed architecture acquires various information from multiple data resources, including route information (speed limits, road grade, position of route markers such as traffic lights and stops signs) from an advanced navigation system, SPaT information from a Dedicated Short Range Communication (DSRC) unit, and target vehicle velocity data from V2V communication or on-board detection system (comprises of radar and camera unit). A simpler but faster variant of LSTM namely Gated Recurrent Unit (GRU) is used to build an Encoder-Decoder (ED) network that provides an estimate of the future target vehicle speed trajectory. Existing literature has shown that GRU's are faster, compact and ideal for short-term predictions (\cite{gao2020short}). The speed predictor is integrated into the Eco-Driving controller to estimate the future speed trajectory of the target vehicle. The optimization is performed in two stages, including a long-term optimization that computes a base policy utilizing the route information from the navigation system available over the entire itinerary, and a receding-horizon optimization that incorporates V2X communications and periodically performs the co-optimization of vehicle velocity and powertrain control over a short horizon. The proposed Eco-Driving controller is evaluated through microscopic simulations.
	
	\section{Mild Hybrid Electric Vehicle  Model}
	\label{model_development}
	A forward-looking parallel mild-Hybrid Electric Vehicle  (HEV) is considered, as shown in \cite{olin2019reducing}. A P0 HEV powertrain with a Belted Starter Generator (BSG) performs torque assist, regenerative braking and start-stop functions over real-world routes. The BSG is connected to the crankshaft of a 1.8L turbocharged gasoline engine and a 48V battery pack.
	
\begin{figure}[t!]
		\begin{center}
			\includegraphics[width=8.4cm]{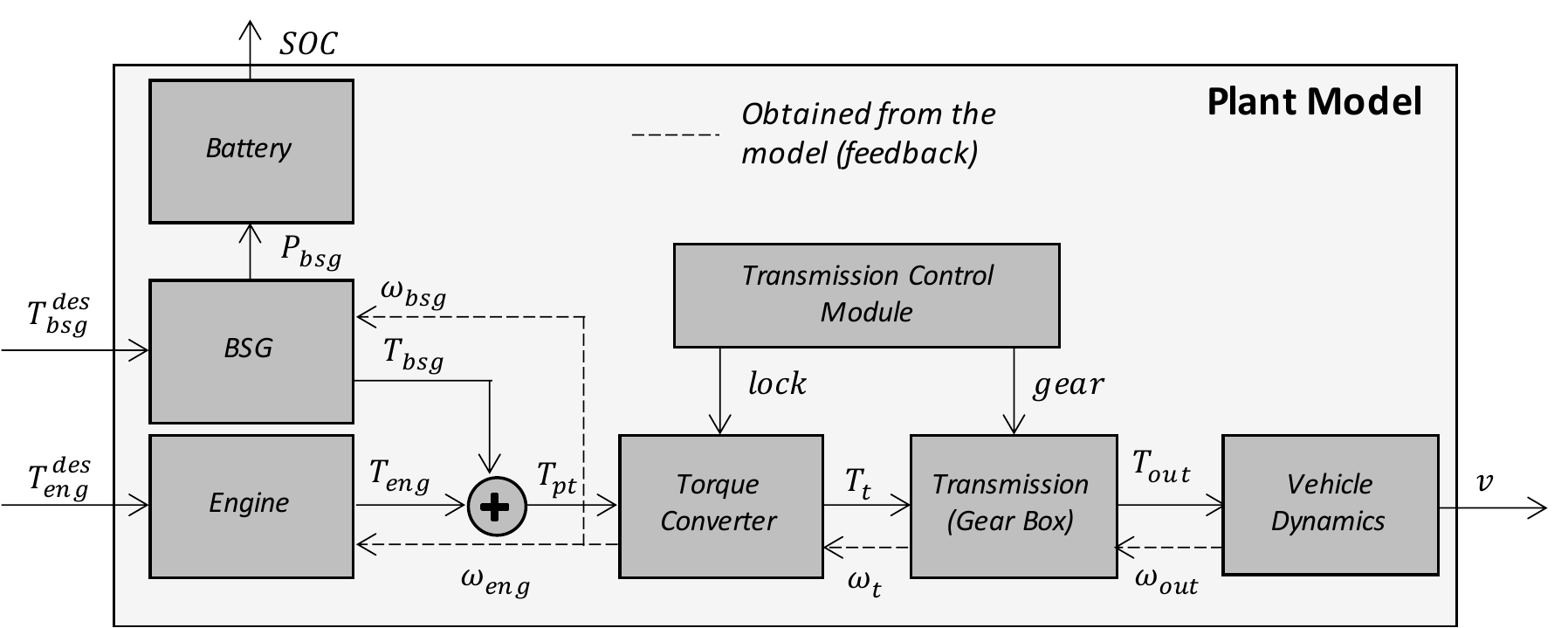}
			\caption{Block Diagram of 48V P0 Mild-Hybrid Drivetrain.}
			\label{fig: plant_model}
		\end{center}
	\end{figure}

	Fig. \ref{fig: plant_model} illustrates the inputs to the Vehicle Dynamics and Powertrain (VD\&PT) model, obtained from a simplified model of the Engine Control Module (ECM) that contains the essential functions to convert the driver’s input (pedal position) to torque commands. The outputs of the ECM, the desired BSG torque ($T^{des}_{bsg}$) and desired engine torque ($T^{des}_{eng}$) are obtained from a production level torque split strategy, which is used as the baseline for fuel economy evaluation.
	
\begin{figure}[t!]
		\centering
		\includegraphics[width=\columnwidth]{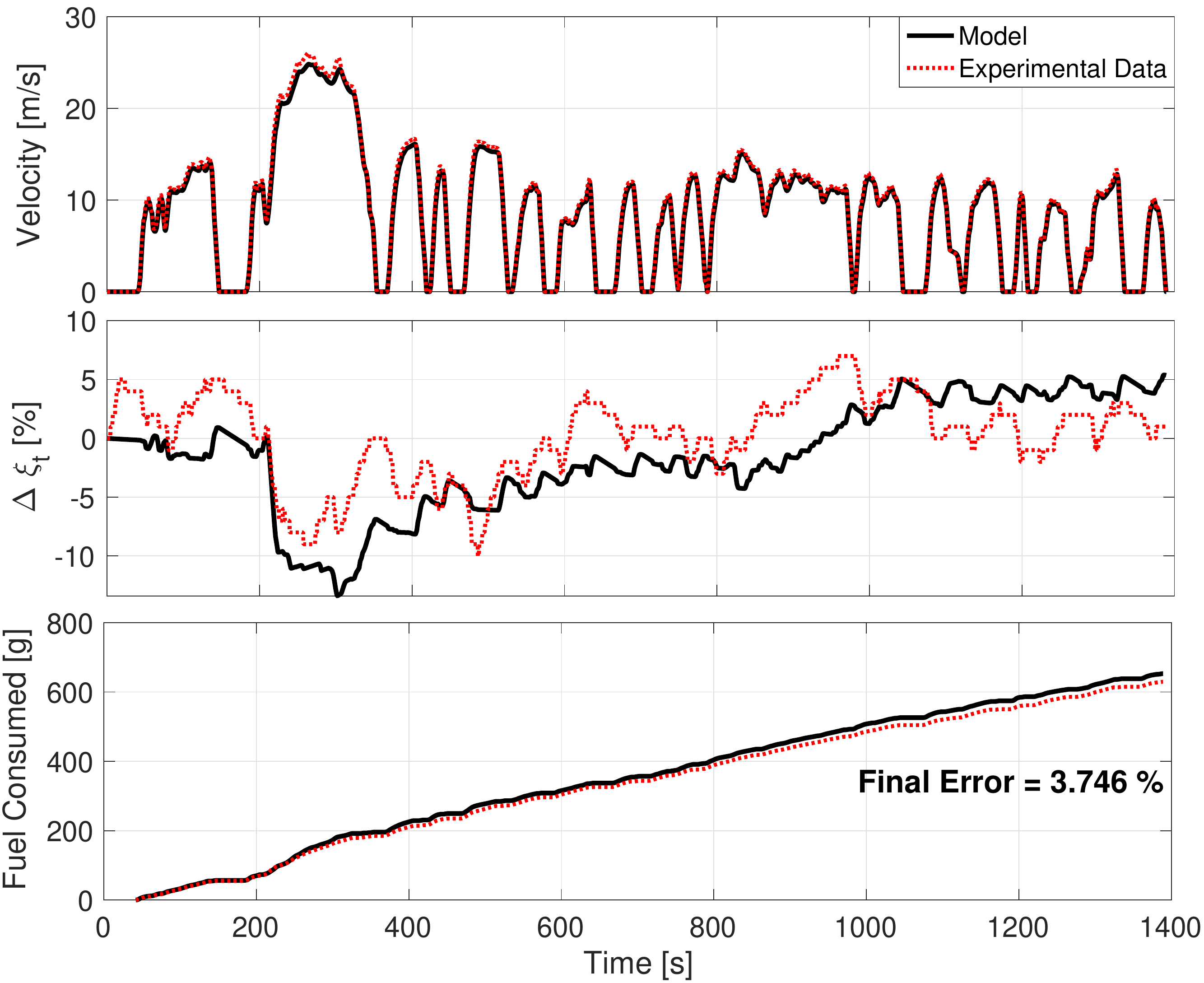}
		\caption{Validation of Vehicle Velocity, SoC ($\xi_t$) and Fuel Consumed over FTP Cycle.}
		\label{fig: veh_vel_soc_fuel_validation_FTP}
	\end{figure}

	The simulator contains low-frequency dynamic models of the powertrain and longitudinal vehicle dynamics. The battery is modeled as a zero-th order equivalent circuit, from which the battery State-of-Charge (SoC) is calculated; the vehicle longitudinal dynamics are described by the road load equation (\cite{guzzella2007vehicle}). Quasi-static models are developed for the engine (fuel and friction maps), BSG, torque converter and transmission (efficiency maps). The model was validated on chassis dynamometer data. Fig. \ref{fig: veh_vel_soc_fuel_validation_FTP} shows a verification over the FTP drive cycle, where the vehicle velocity, battery SoC ($\xi$) and fuel consumption are compared against test data. Despite small mismatches in the battery SoC, attributed to simplification of the 12V electrical system, the fuel consumption is predicted with cumulative error less than 4\%, which is sufficiently accurate for the intended use.
	
	\section{Target Vehicle Velocity Prediction}
	\label{lead_velocity_prediction}
	Prediction of target vehicle velocity is essential for efficiently applying Eco-Driving in urban traffic conditions. Most of the work published to date assume that the target vehicle is either connected to the infrastructure or to the ego vehicle. Since this assumption is rather limiting, the prediction here developed considers that only the current value of the target vehicle velocity is available at each time step. This represents the most general case of an ego vehicle equipped with an object detection system. 
	
	The short term planning of a typical human driver with respect to traffic conditions is generally represented as a time-varying, nonlinear time-series in terms of speed and acceleration (\cite{toledo2007integrated, gupta2019enhanced}). Despite the flexibility and power of Deep Neural Networks (DNNs), they are not ideal for sequences, since they require the dimensionality of input and output to be known and fixed (\cite{sutskever2014sequence}). In this work, the time-dependent nature of the driver behavior is captured by using sequence-to-sequence modeling via a Gated Recurrent Unit Encoder-Decoder.
	
	\subsection{Gated Recurrent Unit Encoder-Decoder Architecture}  \label{GRU_ED}
	The concept of Gated Recurrent Unit (GRU) was originally proposed by \cite{chung2014empirical} to overcome the vanishing gradient problem in Vanilla RNNs (\cite{pascanu2013difficulty}) and make each recurrent unit adaptively capture temporal dependencies of sequential data. A GRU network has two gating units: a reset gate $r_t$ and an update gate $z_t$ that modulate the flow of information inside the unit. The reset gate determines how much of the past information to forget, and the update gate determines how much of the past information needs to be passed along to the future. Even though GRUs do not explicitly contain separate memory cells as in LSTM (\cite{hochreiter1997long}), the memory is introduced in the network by the hidden state vector $h_t$, which is unique for each input sequence. This makes the GRU relatively fast and compact, suitable for short-term predictions (\cite{gao2020short}).
	
\begin{figure}[t!]
		\centering
		\includegraphics[width=\columnwidth]{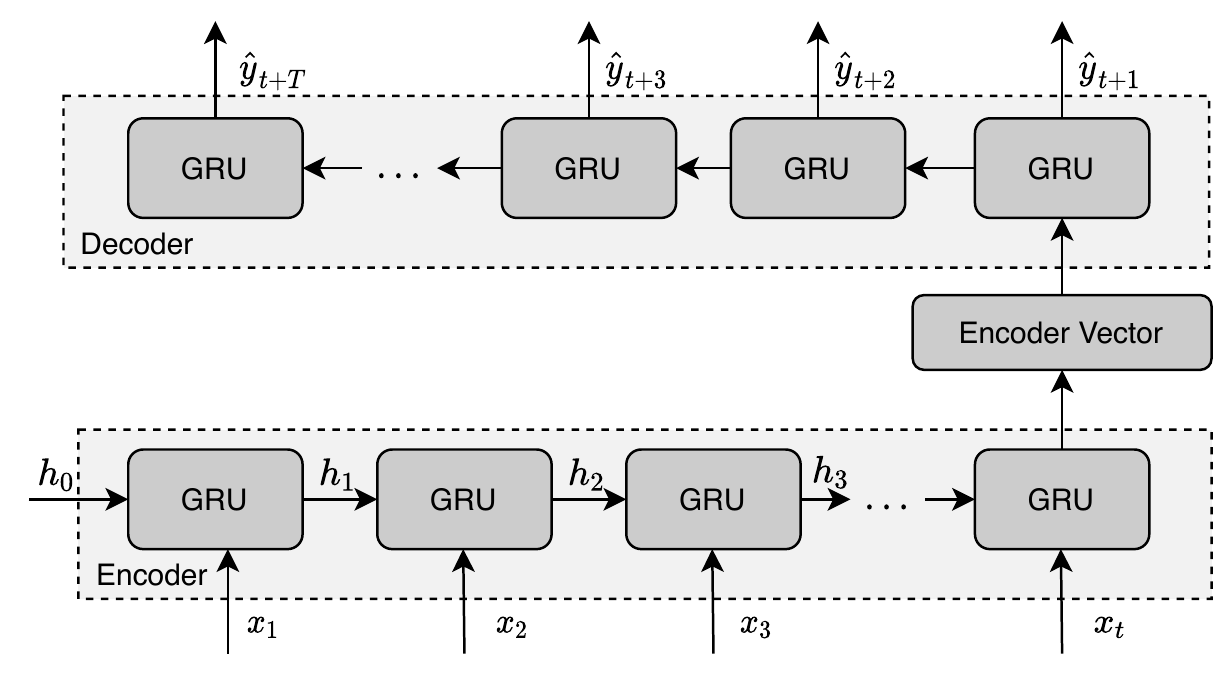}
		\caption{Block Diagram of GRU-ED Architecture.}
		\label{fig: GRU_ED_architecture}
	\end{figure}

	In this work, a GRU-Encoder-Decoder (GRU-ED) architecture has been used to perform sequence-to-sequence prediction of the target vehicle velocity. In literature, GRU-ED has been extensively used in machine translation tasks (\cite{cho2014learning}). GRU-ED is a sequentially connected 3-layer arrangement comprising of an encoder, hidden and decoder layer, as shown in Fig. \ref{fig: GRU_ED_architecture}. The encoder layer contains GRU cells that read input sequences one time step at a time to obtain a large fixed-dimensional encoder vector that intrinsically learns the representation in the input sequence. The encoded vector is then fed as input to another sequence of GRU cells contained in the Decoder layer that generates the predicted sequence.
	
	Given an input sequence, $\mathrm{X}=(x_1,x_2,\cdots,x_t)$, where $x_i\in\mathbb{R}^d$ and $d$ is the feature length of each time step, the GRU encoder computes the sequence of hidden state: $h_1,h_2,\cdots,h_t$ (Fig. \ref{fig: GRU_ED_architecture}) using the equations (\cite{zhang2017gru}):
	\begin{subequations}
		\label{eqn: GRU_encoder}
		\begin{align}
			z_t & = \sigma(W_{xz}x_t+U_{hz}h_{t-1})\\
			r_t & = \sigma(W_{xr}x_t+U_{hr}h_{t-1})\\
			\tilde{h}_t & = \tanh\left(W_{xh}x_t+U_{rh}(r_t\otimes h_{t-1})\right)\\
			h_t & =(1-z_t)\otimes h_{t-1}+z_t\otimes \tilde{h}_t
		\end{align}
	\end{subequations}
	where $\sigma$ is the sigmoid function and $\otimes$ is an element-wise multiplication operator; $z_t, r_t, \tilde{h}_t$ are the update gate, reset gate and candidate activation respectively. $W_{xz}, W_{xr}, W_{xh}, U_{hz}, U_{hr}, U_{rh}$ are related weight matrices. The decoder then similarly predicts an output sequence $\mathrm{Y}=(\hat y_{t+1},\hat y_{t+2},\cdots,\hat y_{t+T})$, where $\hat y_i\in\mathbb{R}$ is the predicted velocity.
	
	\subsection{Driving Features} \label{driving_features}
\begin{figure}[t!]
		\centering
		\includegraphics[width=\columnwidth]{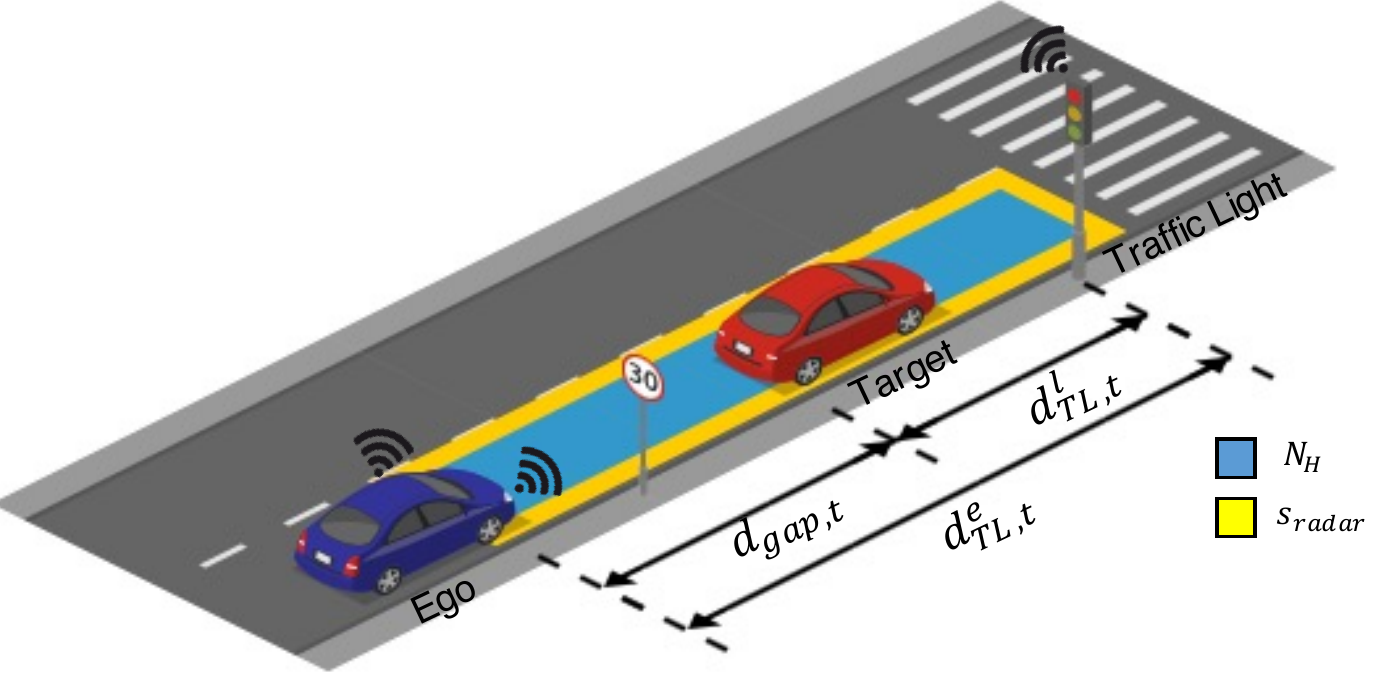}
		\caption{Schematic of the ego vehicle approaching a signalized intersection while following an unconnected target vehicle.}
		\label{fig: ego_lead_vehicle}
	\end{figure}

	The crucial step in developing a vehicle speed prediction model is the selection of a set of input features that are representative of real-world driving. In this work, the input features were selected to be easily measured from ego vehicle's on-board detection system, without requiring the target vehicle to broadcast its position and velocity. However, it is assumed that V2I communication exists, and more specifically that signalized intersections can broadcast SPaT messages to the ego vehicle within a predetermined transmission range. The longitudinal motion of any vehicle on a straight and flat road over the next time step $\Delta t$ can be determined by the velocity $v_t$ and acceleration $a_t$ at the current time $t$:
	\begin{equation}
		\label{eqn: longitudinal_dynamics}
		\begin{aligned}
			v_{t+1}=v_t+a_t\Delta t.
		\end{aligned}
	\end{equation}
	This means that the pair $[v_t, a_t]$ is used to develop non-linear car-following longitudinal models (\cite{gupta2019enhanced, rajakumar2020benchmarking}). Furthermore, it is necessary to predict the target vehicle's behavior near a signalized intersection, as shown in \cite{gupta2020estimation}. This implies that the driver behavior near an intersection can be expressed as a non-linear function of velocity, acceleration and distance to the traffic light ($d_{TL,t}^{l}$). Finally, the feature set $\mathcal{F}=\{v_{t}^l, a_{t}^l, d_{TL,t}^{l}\}$ can comprehensively represent human driving on a straight road and near signalized intersections. Note that the assumption of a human driving and no connectivity in the target vehicle does not allow the ego vehicle to incorporate a car-following behavior in the feature set, for example as done by \cite{hyeon2021forecasting}. However, even without V2V connectivity, the distance of the target vehicle to the traffic light $d_{TL,t}^l$ can be evaluated from the assumption of the ego vehicle receiving V2I communication within a communication range, as shown in Fig. \ref{fig: ego_lead_vehicle}. Outside this range, $d_{TL,t}^l$ is arbitrarily set to 500m:
	\begin{equation}
		\label{eqn: lead_position}
		\begin{aligned}
			d_{TL,t}^l=\begin{cases}
				d_{TL,t}^e-d_{gap,t}, \quad d_{TL,t}^e<d_{DSRC}\\
				500m, \quad \text{otherwise}
			\end{cases}
		\end{aligned}
	\end{equation}
	where $d_{TL,t}^e$ refers to the distance of the ego vehicle to the intersection obtained via V2I communication (position of upcoming intersection is broadcasted by the V2I modem); $d_{g,t}$ refers to the relative distance between the ego and target vehicle, obtained via on-board radar unit.
	
	\subsection{Methodology}
	A comprehensive vehicle trajectory dataset, Next Generation Simulation (NGSIM), was collected by U.S. Federal Highway Administration (FHWA) in 2005 and is widely used in transportation research, especially in traffic flow analysis and modelling, traffic-related estimation and prediction, and vehicular ad hoc network-related studies (\cite{kovvali2007video}). In this paper, a subset of the dataset is adopted to train the GRU-ED network.
	
	The GRU-ED network is trained over approximately 18 hours of NGSIM driving data using TensorFlow and the Keras API package. To access the learning performance of the model and its ability to generalize over different drivers, the considered data-set is split into 70\% for training the GRU-ED and 30\% to validate and test the trained model. For better generalization over the different features, the feature set $\mathcal{F}$ is normalized between 0 and 1 using the sci-kit learn toolbox in Python. Custom functions are designed to split the data into a historical window and prediction window of time duration  $T_h$ and $T_p$ respectively during the training phase. To evaluate the performance of the prediction model, Root Mean Square Error (RMSE) is computed between the predicted speed ($\hat y_t$) and the actual speed ($ y_t$):
	\begin{equation}
		\label{eqn: RMSE}
		\text{RMSE} = \sqrt{\frac{1}{N}\sum_{i=1}^{N}(y_t-\hat y_t)^2}
	\end{equation}
	
	\subsection{Evaluation}
	To analyze the effect of both $T_h$ and $T_p$ on the accuracy of prediction, a sensitivity study was performed where the GRU-ED network is trained and tested over several combinations of $[T_h,T_p]$ pairs, as shown in Table 1. 
	
	For a given $T_h$, increasing the length of prediction window increases the RMSE. For instance, the pair $[T_h,T_p]=[5s,5s]$ produces the lowest RMSE and as the length of prediction window increases for fixed $T_h$, the RMSE increases. However, for a given $T_p$, increasing the length of $T_h$ decreases the RMSE but the difference is not very large. This signifies that the network's accuracy is more sensitive to the length of prediction window but less sensitive to the length of historical window.
	
	\begin{table}[ht!]
		\label{tab: RMSE_T_h_T_p}
		\caption{RMSE between predicted and actual velocity for different $T_h$ and $T_p$}
		\centering
		\begin{tabular}{|c|c|c|c|c|c|c|}
			\hline
			\begin{tabular}[c]{@{}c@{}}{[}$T_h,T_p${]}\\ {[}s,s{]}\end{tabular} & {[}5,5{]} & {[}5,10{]} & {[}5,20{]} & {[}10,10{]} & {[}10,20{]} & {[}20,20{]} \\ \hline
			\begin{tabular}[c]{@{}c@{}}RMSE\\ {[}m/s{]}\end{tabular}            & 0.77          &  1.85          &  3.45          &1.83             &  3.44           &3.36             \\ \hline
		\end{tabular}
	\end{table}
	
\begin{figure}[t!]
		\centering
		\includegraphics[width=\columnwidth]{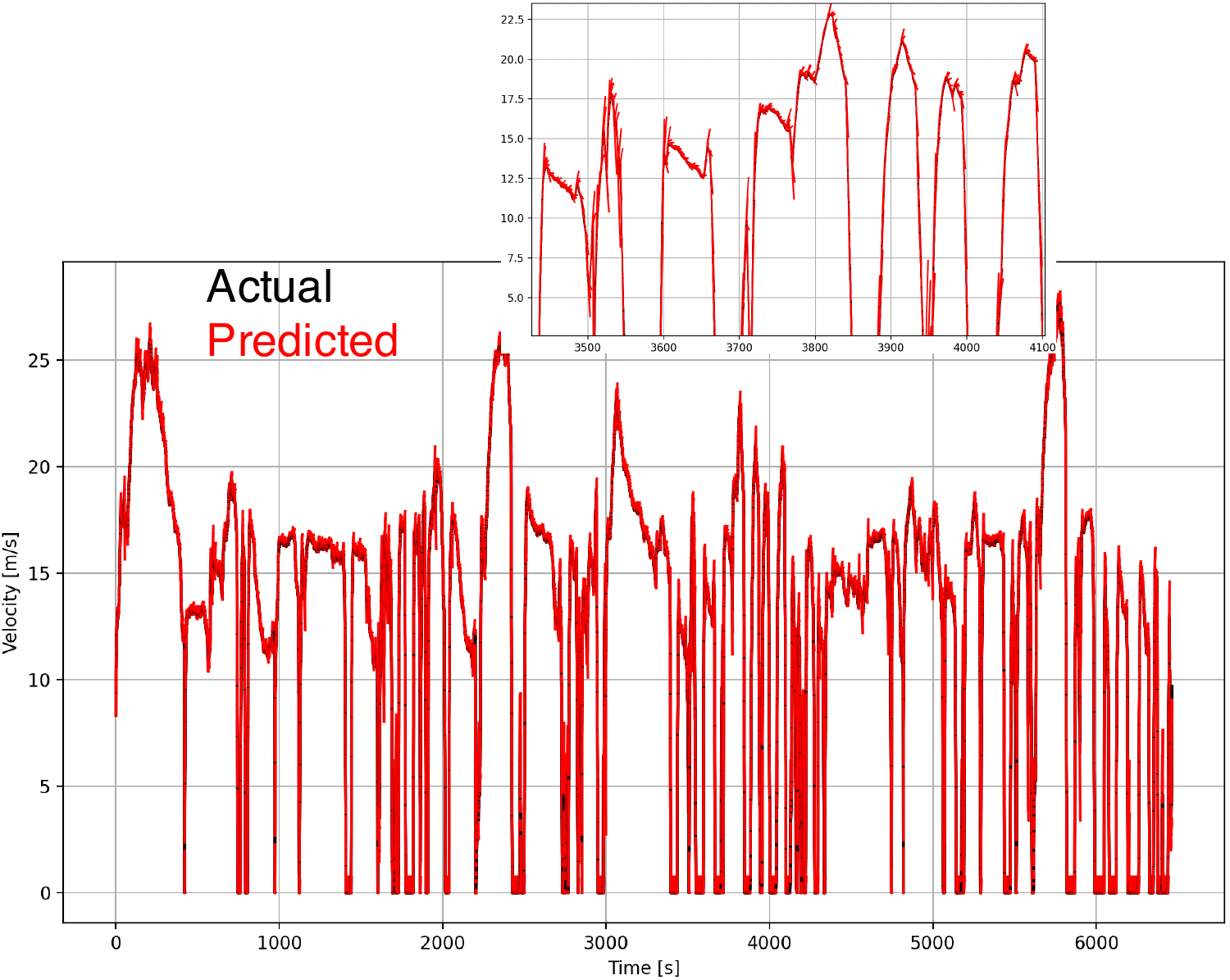}
		\caption{Comparison between the Predicted and Actual Velocity Trajectory.}
		\label{fig: prediction_results_5s}
	\end{figure}

\begin{figure}[t!]
		\centering
		\includegraphics[width=\columnwidth]{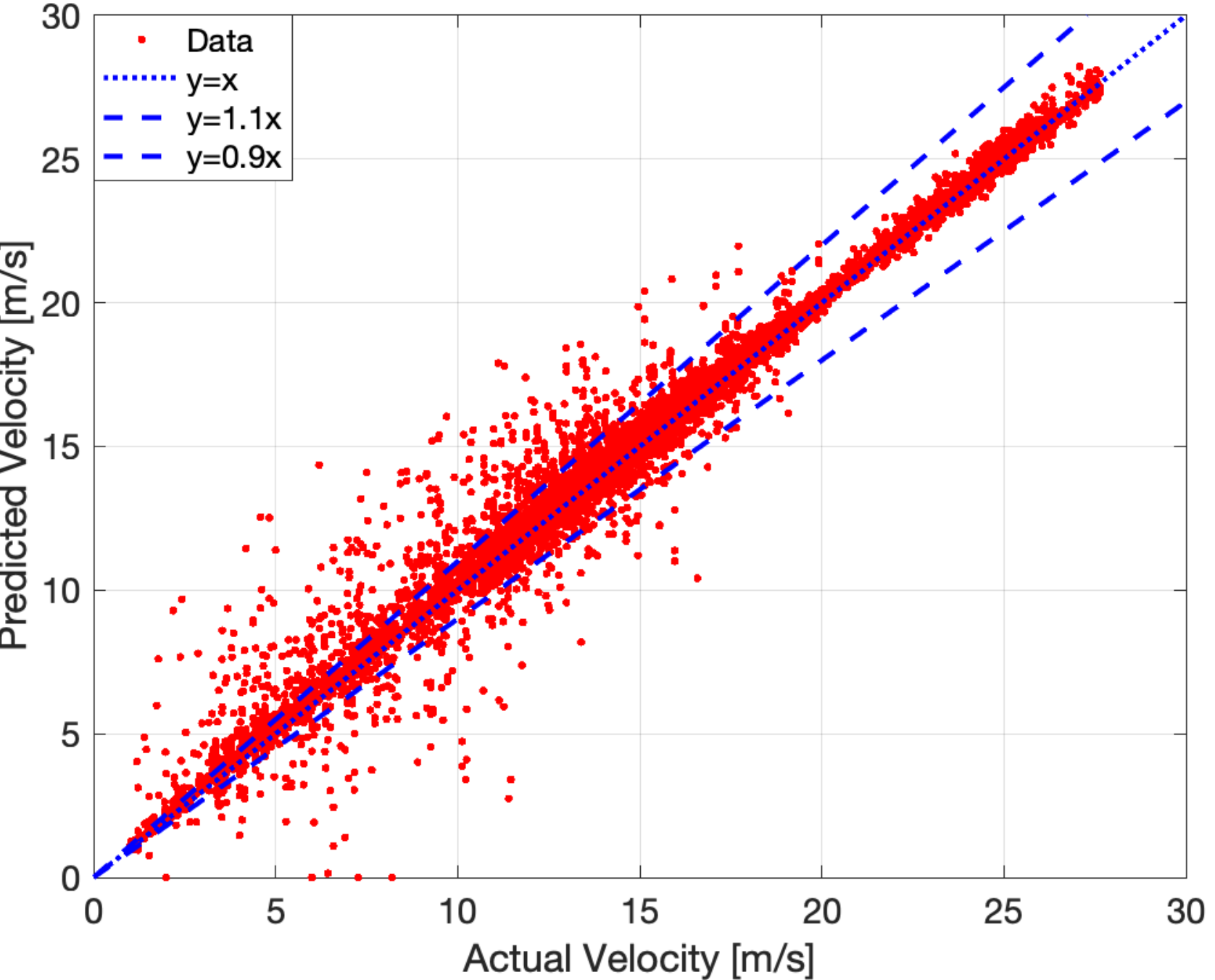}
		\caption{Correlation Plot for $T_h=5s$ and $T_p=5s$.}
		\label{fig: corr_plot_5}
	\end{figure}

	Fig. \ref{fig: prediction_results_5s} and Fig. \ref{fig: corr_plot_5} show the actual and predicted velocity comparison and the correlation plot respectively for $[T_h,T_p]=[5s,5s]$. Clearly, the predictor is able to forecast most of the acceleration/deceleration/cruise events, but it diverges for small durations at points where the vehicle abruptly switches from an acceleration to a deceleration maneuver (or vice-versa). This is understandable, since the velocity predictor relies only on the current values of velocity and acceleration, hence large variations in magnitude and sign of the vehicle acceleration (for example occurring during a rapid transition from acceleration to braking) are impossible to predict with the available features. Nonetheless, the network converges to the actual velocity within the span of few seconds.
	
	\section{Integration with Eco-Driving Algorithm}
	\label{prbm_formulation}
	The Eco-Driving algorithm implemented in this study is based on an Optimal Control Problem (OCP) formulated in spatial domain that minimizes fuel consumption over an entire route of N steps:
	\begin{equation}
		\label{eq: ed_ocp}
		\min_{u_s} \sum_{s=1}^{s=N}{\left(\gamma \cdot \dot{m}_{\mathrm{f},s} +(1-\gamma)\right)\cdot \Delta t_s}
	\end{equation}
	where $s$ is the discrete position, $u_s$ is the control input, $\gamma$ is a weighing factor between the fuel consumption and the travel time, and $\dot{m}_{\text{f},s}$ is the fuel flow rate. $\Delta t_s (:=\frac{\Delta d_s}{\bar{v}_{s}})$ is the travel time per step, $\Delta d_s$ is the distance step (i.e. $\Delta d_s=d_{s+1}-d_s$) calculated from the distance traveled ($d_s$) along the route at position $s$, and $\bar{v}_{s}=\frac{v_{s}+v_{s+1}}{2}$ is the average velocity. The state and action space are subject to a set of constraints, as described in \cite{olin2019reducing}.
	
	To incorporate time-based information such as SPaT and presence of other vehicles, time is added as an additional state variable in the OCP formulated in the spatial domain. In previous work, \cite{zhu2021gpu} formulated a 3 state OCP with time as one of the states to incorporate the SPaT information into the Eco-Driving controller, assuming a traffic-free environment. However, the effectiveness of this controller would reduce in presence of a target vehicle, particularly near signalized intersections. In this work, the target vehicle velocity predicted by the GRU-ED network is integrated with the Eco-Driving OCP and solved over a receding horizon of $N_H (N_H<<<N)$ steps to mitigate the aforementioned issue.
	
	One way to incorporate target vehicle velocity information into the Eco-Driving controller would be to impose the predicted target vehicle velocity as a state constraint in the OCP, however this would likely result in an overly conservative design. %For instance, it is not necessary for the ego vehicle to decelerate if the target vehicle at a large separation distance is decelerating. 
	For this reason, the target vehicle velocity trajectory ($\hat v^l_{s}$) predicted by the GRU-ED network is utilized within the Eco-driving controller to determine the relative distance $\hat d_{gap,s}$ between the lead and the ego vehicle, whose velocity ($v^e_{s}$) is computed as the solution of the receding horizon OCP.
	
\begin{figure}[t!]
		\centering
		\includegraphics[width=\columnwidth]{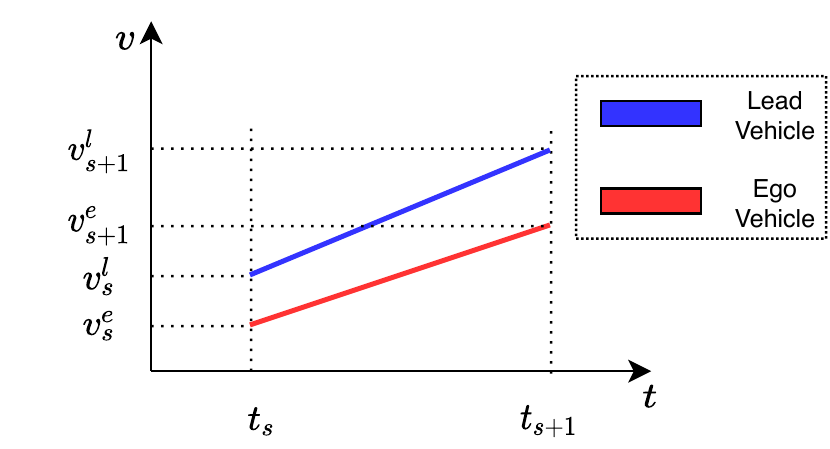}
		\caption{Ego and Target Vehicle Velocity over $\Delta d_s$.}
		\label{fig: lead_constraint}
	\end{figure}

	To compute the relative distance $\hat d_{gap,s}$, let us consider that at a time $t_s$, the controller has the information about its current velocity $v^e_s$, current target vehicle velocity $v^l_s$ and the current distance gap $d_{gap,s}$ available from the on-board detection system. Further let us assume that the ego vehicle takes $\Delta t_s$ time to travel a distance step $\Delta d_s$ and reaches a velocity $v^e_{s+1}$ as shown in Fig. \ref{fig: lead_constraint}. The target vehicle velocity at time $t_{s+1}$ is obtained by interpolating the time-varying predicted velocity trajectory obtained from by GRU-ED network. The difference in the area between the ego and lead velocity-time curve would denote the change in relative distance $\Delta d_{gap,s}$  over the next distance step. At a given position $k$, the dynamic model for $d_{gap,k}, \forall k=s,\cdots,s+N_H-1$ can be derived as:
	\begin{subequations}
		\label{eq: realtive_gap_model}
		\begin{align}
			\hat d_{gap,k+1}&=\hat d_{gap,k} + (\bar{\hat v}_k^l-\bar{v}_k^e)\Delta t_k\\
			&=\hat d_{gap,k} + (\bar{\hat v}_k^l-\bar{v}_k^e)\frac{\Delta d_k}{\bar{v}^e_k}\\
			&=\hat d_{gap,k} + \left(\frac{\bar{\hat v}_k^l}{\bar{v}_k^e}-1\right){\Delta d_k}   \label{eq: realtive_gap_model_c}		
		\end{align}
	\end{subequations}
	
	Eqn. \ref{eq: realtive_gap_model_c} represents the relative distance between the lead and ego vehicle at any point in a $N_H$ horizon, which varies as a function of the ratio of average lead and ego vehicle velocity evaluated over a distance step. If $\bar{\hat v}_k^l>\bar{ v}_k^e$, the target vehicle is faster than ego vehicle and the $\hat d_{gap,{k+1}}>\hat d_{gap,{k}}$. On the other hand, if  $\bar{\hat v}_k^l<\bar{ v}_k^e$, the target vehicle is slower than ego vehicle and the $\hat d_{gap,{k+1}}<\hat d_{gap,{k}}$. This logic can be used by the Eco-Driving controller to modulate the ego vehicle velocity such that the $\hat d_{gap,{k}}>d_{safe,s}$ which is defined as the safe car-following distance (\cite{brackstone1999car}):
	\begin{equation}
		\label{eq: car-following_model}
		d_{safe,s} = d_0 + v^e_{s+1} T_{\text{gap}} +\frac{v^e_{s+1} \Delta v_{s+1}}{2\sqrt{a_{\text{max}}b_{max}}}
	\end{equation}
	where $d_0$ refers to the safe distance gap at stand still, $T_{gap}$ refers to the time gap, $\Delta v_{s+1}$ refers to the relative velocity between target and ego vehicle, $a_{\text{max}}$ and $b_{max}$ refers to the maximum acceleration and deceleration respectively. Assume that the discretized state dynamics for the Eco-Driving problem has the following form:
	\begin{equation}
		x_{s+1}=f_s(x_s,u_s), \quad s=1,\cdots, N-1.
	\end{equation}
	where $x_s=[v_s,\xi_s,t_s]^\mathsf{T}\in\mathcal{X}_s\subseteq\mathbb{R}^p$ and $u_s=[T_{\mathrm{eng},s}, T_{\mathrm{bsg},s}]^\mathsf{T}\in\mathcal{U}_s\subseteq\mathbb{R}^q$ are the state and control actions respectively. In this work, the state variables are the vehicle velocity $(v_s)$, battery SoC $(\xi_s)$ and travel time $(t_s)$. The control actions are the engine torque $(T_{\mathrm{eng},s})$ and BSG torque $(T_{\mathrm{bsg},s})$. The equations describing the state dynamics $f_s(x_s, u_s)$ have been derived in prior work (\cite{gupta2019thesis}).
	
	As mentioned above, the Eco-Driving optimization problem is formulated as a receding horizon optimal control problem where the full route of $N$ steps is solved over a reduced horizon $N_H (<<N)$. At a given position $s = 1, \dots, N-N_H$, the optimization problem is formulated as:
	\begin{equation}
		\label{eq: ed_ocp_N_H_horizon}
		\begin{gathered}
			\mathcal{J}^*(x_s)= \min_{\left\lbrace \mu_k \right\rbrace_{k=s}^{s+N_H-1}}\sum_{k=s}^{s+N_{H}-1} c_\mathrm{T}(x_{s+N_H}) + c(x_k,\mu_{k}(x_{k})), \\
			c(x_k,\mu_{k}(x_{k}))=\left(\gamma \cdot \dot{m}_{\mathrm{f},k}(x_k,\mu_{k}(x_{k})) +(1-\gamma)\right)\cdot \Delta t_k
		\end{gathered}
	\end{equation}
	where $\mu_k:\mathcal{X}\rightarrow\mathcal{U}$ is the admissible control policy of the controller at the step $k$ in the prediction horizon; $c:\mathcal{X}\times\mathcal{U}\rightarrow\mathbb{R}$ is the stage cost function defined as the weighted average of the fuel consumption and travel time; $c_{\mathrm{T}}:\mathcal{X}\rightarrow\mathbb{R}$ is the terminal cost function. The state space and action space are subject to following constraints: $\forall s = 1,\dots,N-N_H$, $\forall k=s,\dots,s+N_H$:
	\begin{subequations}
		\label{eq: constraints_N_H_horizon}
		\begin{align}
			v_k&\in[v_k^{\mathrm{min}},v_k^{\mathrm{max}}],\\
			\xi_k&\in[\xi_k^{\mathrm{min}},\xi_k^{\mathrm{max}}],\\
			t_k&\in\mathcal{T}_{\mathrm{G},k},\\
			a_k&\in[a^{\mathrm{min}},a^{\mathrm{max}}],\\
			\hat d_{gap,k}&\in [d_{\text{safe}},d_{\text{radar}}]\\
			T_{\mathrm{eng},k}&\in[T_{\mathrm{eng}}^{\mathrm{min}}(v_k),T_{\mathrm{eng}}^{\mathrm{max}}(v_k)],\\
			T_{\mathrm{bsg},k}&\in[T_{\mathrm{bsg}}^{\mathrm{min}}(v_k),T_{\mathrm{bsg}}^{\mathrm{max}}(v_k)],
		\end{align}
	\end{subequations}
	where $v^{\mathrm{min}}_s, v^{\mathrm{max}}_s$ are the minimum and maximum route speed limits respectively, $\xi_s^{\mathrm{min}}, \xi_s^{\mathrm{max}}$ are the static limits applied on battery SoC, $a^{\mathrm{min}}, a^{\mathrm{max}}$ represent the limits imposed on the acceleration for comfort, $d_{\text{safe}}$ is obtained from Eqn. \ref{eq: car-following_model}, $d_{radar}$ refers to the radar range (assumed to be 250m), $T_{\mathrm{eng}}^{\mathrm{min}}(v_s),T_{\mathrm{eng}}^{\mathrm{max}}(v_s)$ are the minimum and maximum engine torque limits, and $T_{\mathrm{bsg}}^{\mathrm{min}}(v_s),T_{\mathrm{bsg}}^{\mathrm{max}}(v_s)$ are the minimum and maximum BSG torque limits, respectively. To ensure SoC neutrality over the entire itinerary, a terminal constraint $\xi_1=\xi_N$ is applied to the battery. $\mathcal{T}_{\mathrm{G},s}$ represents the feasible set of travel time for passing-at-green at signalized intersections (\cite{zhu2021gpu}).
	
	The receding horizon OCP is solved using Approximate Dynamic Programming (ADP), such that an approximate terminal cost used in the receding horizon OCP (named \say{Short-term Optimization} in Fig. \ref{fig: eco-drving-architecture}) is obtained from the offline solution of a full-route optimization under partial information (termed \say{Long-term Optimization}). 
	
	\section{Simulation and Evaluation of Results}
\begin{figure}[t!]
		\centering
		\includegraphics[width=\columnwidth]{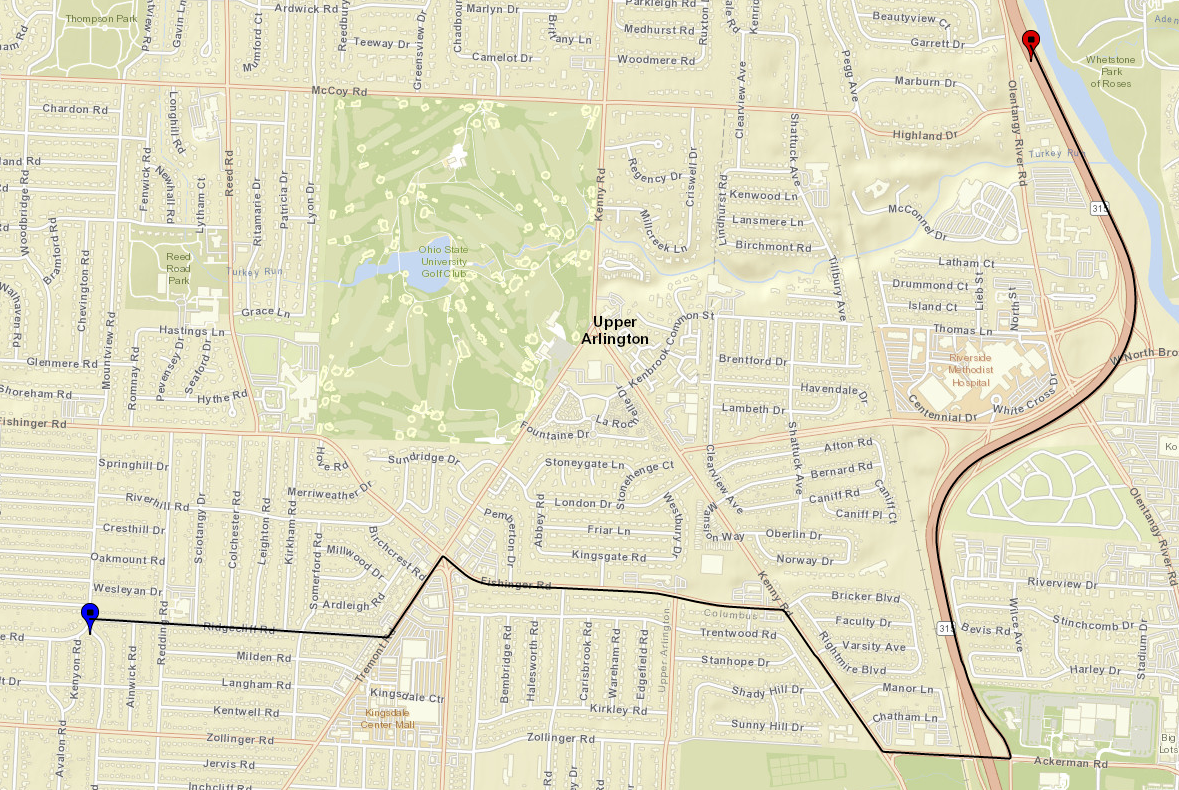}
		\caption{Selected route for simulation and validation on OpenStreetMaps (Route 19). Blue and red marker denote the start and destination respectively.}
		\label{fig: Route19}
	\end{figure}

	To illustrate the proposed approach, a mixed-urban route in Columbus, OH (USA) was been selected for the simulation and analysis. As shown in Fig. \ref{fig: Route19}, the route is 7km in length and comprises 5 traffic lights and 2 stop signs (start and end of the trip).
	
	To simulate the presence of a target vehicle, a time-varying vehicle velocity profile is generated using Simulation of Urban MObility (SUMO) for a given departure time. 
	The vehicle velocity profile generated in SUMO is used as the source for the velocity predictor integrated into the Eco-Driving controller. The velocity profile is segmented into historical data using a moving window of 10s, then fed to the trained GRU-ED network to predict the target vehicle velocity within the receding horizon OCP that determines the optimal speed and battery SoC profiles for the ego vehicle (Eqn. \ref{eq: ed_ocp_N_H_horizon} and \ref{eq: constraints_N_H_horizon}). Note that this framework is reasonable for this study, since the vehicle motion in the simulation is limited to a single lane and the SPaT sequences obtained from SUMO are post-processed to be consistent with the departure time of the target vehicle.
	
\begin{figure}[t!]
		\centering
		\includegraphics[width=\columnwidth]{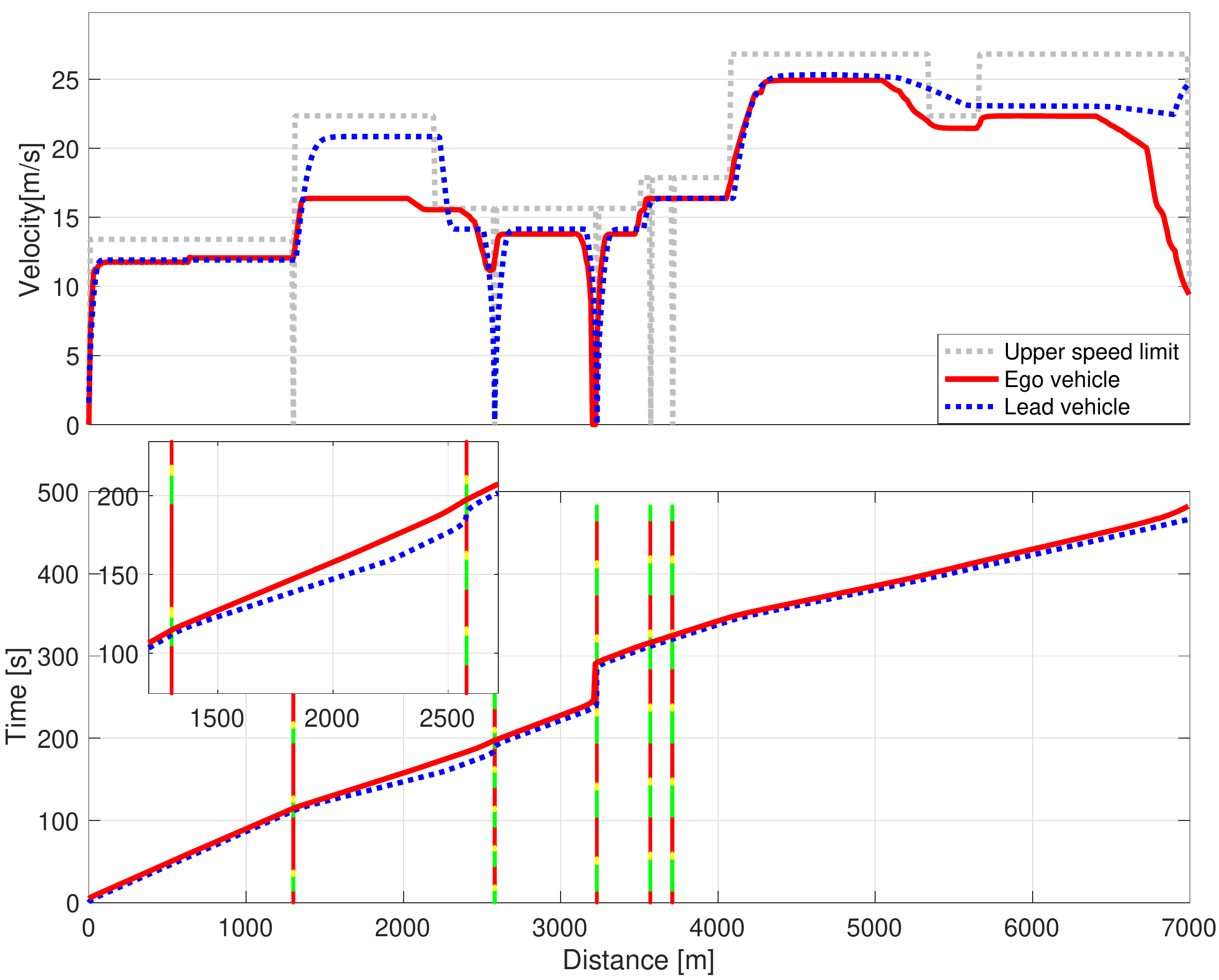}
		\caption{Speed trajectory and Time-Space plot showing the eco-driving manuever in presence of V2I communication and human-driven target vehicle over Route 19.}
		\label{fig: time_space_plot_dist}
	\end{figure}

\begin{figure}[t!]
		\centering
		\includegraphics[width=\columnwidth]{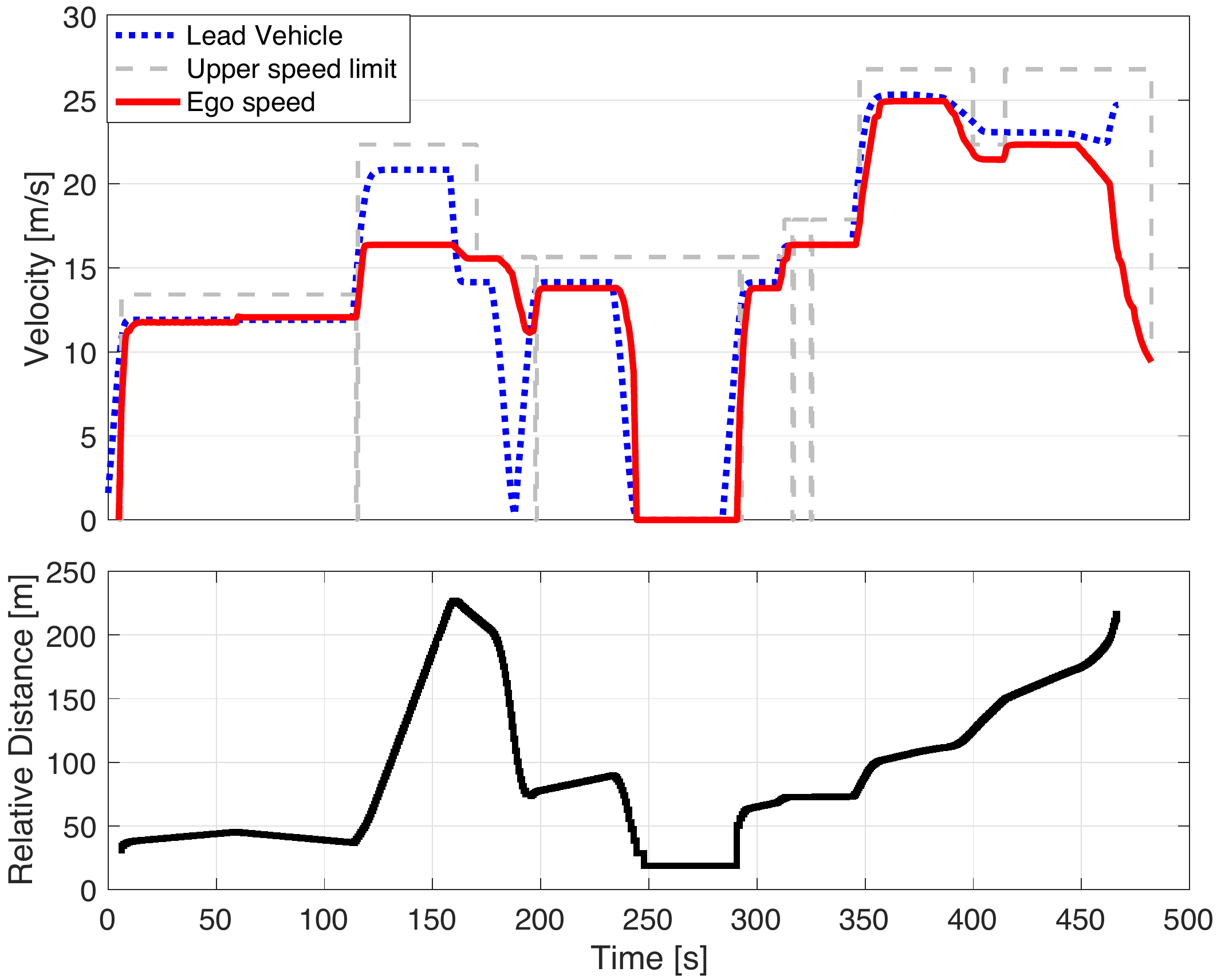}
		\caption{Speed trajectory and Relative Distance to the Target Vehicle over Route 19.}
		\label{fig: rel_dist}
	\end{figure}

	Fig. \ref{fig: time_space_plot_dist}  shows the velocity trajectory and time-space plot of the lead and ego vehicle over the route shown in Fig. \ref{fig: Route19}. The ego vehicle is able to pass all the traffic lights, except the second, at a green phase even in the presence of a target vehicle. Within the V2I communication range of the first, fourth and fifth traffic lights (located at approximately 1300, 3600, 3800m), the Eco-Driving controller is aware of the time remaining in the green phase. This information, combined with the GRU-ED network predicting the target vehicle to cruise (based on the historical data and distance to traffic light), is exploited by the Eco-Driving controller to either hold the ego vehicle speed or accelerate to pass the signalized intersections during the green phase. This is also evident from the Fig. \ref{fig: rel_dist}, where the relative distance either remains the same or decreases.
	
	Moreover, while approaching the second traffic light, the Eco-Driving controller knows \say{a priori} that the time remaining in the red phase is small and the GRU-ED network predicts the target vehicle will decelerate (based on the initial deceleration of the target vehicle while approaching the intersection). Using this information, the controller lowers the ego vehicle's speed to allow the signal to change to green, while at the same time avoiding a collision with the target vehicle. At the third traffic light, since the time remaining in the red phase is large, the Eco-Driving controller decides to stop, to avoid a collision with the target vehicle at the intersection. It should be noted that even during stand-still, the ego vehicle maintains a safe gap from the target vehicle.
	
	To evaluate the benefits of the integration of GRU-ED network with the Eco-Driving controller in terms of Fuel Consumption (FC) and Travel Time (TT), the simulation results were compared against a baseline case where the target vehicle velocity is assumed to be constant over the $N_H$ horizon and equal to the value detected at each time step. Table 2 compares the average FC and TT obtained with a constant velocity predictor and the GRU-ED network predictor while driving over the Route 19 for 5 different target vehicle velocity and signal phase departure times. Results show that the average TT and FC evaluated over the 5 scenarios for constant velocity case is approximately 7\% and 3\% respectively higher than the GRU-ED Network case. This indicates that the GRU-ED velocity forecast provides more opportunities for regen and minimizes unnecessary braking thereby improving the overall fuel economy in all aspect of the optimization. Further, the forecast of target vehicle velocity over the prediction horizon allows the ego vehicle to pass-at-green at most of the intersections when compared against the baseline case reducing overall travel time.
	
	\begin{table}[t!]
		\label{tab: comparison_fc_tt}
		\caption{Comparison of FC and TT with and w/o a lead velocity predictor over Route 19}
		\centering
		\begin{tabular}{|c|c|c|}
			\hline
			Velocity Prediction Method & \begin{tabular}[c]{@{}c@{}}Fuel Consumption\\ {[}g{]}\end{tabular} & \begin{tabular}[c]{@{}c@{}}Tavel Time \\ {[}s{]}\end{tabular} \\ \hline
			Constant Velocity          &                               277 &              520                  \\ \hline
			GRU-ED Network        &            268                                                      &                 486                                           \\ \hline
		\end{tabular}
	\end{table}

	\section{Conclusions}
	In this work, an Eco-Driving strategy formulated for a connected and automated mild-HEV with an on-board detection system, V2I communication and longitudinal automation was augmented with a prediction-based target vehicle velocity estimator. A sequence-to-sequence velocity predictor, based upon a Gated Recurrent Encoder-Decoder (GRU-ED) network was trained to forecast the target vehicle velocity trajectory based upon historic data, and was then integrated with the Eco-Driving controller by designing a dynamic model of the relative distance between the target and the ego vehicle. Compared to previous studies where the target vehicle is assumed to be fully connected, the methodology proposed in this paper is applicable to the more general case of a target vehicle with partial or no automation, and without the ability to broadcast velocity and acceleration.
	
	The Eco-Driving problem was cast as a receding horizon optimal control problem with three states, namely vehicle velocity, battery SoC and travel time. The problem was solved using Approximate Dynamic Programming and implemented in a Model Predictive Control (MPC) framework. It is worth mentioning that the formulation presented in this paper integrates all elements of the Eco-Driving problem (speed planning and powertrain set-points optimization) into a single controller. Simulation results indicate that the Eco-Driving controller is able to intelligently pass the signalized intersection in green phase, by forecasting the target vehicle velocity. Future work will address the development of a co-simulation platform where the optimization is performed in a parallel implementation developed with CUDA programming on a GPU to reduce the computational requirement from inclusion of additional states and allow to perform Eco-Driving in presence of multiple surrounding vehicles in a multi-lane scenario.
	
	\begin{ack}
		The authors acknowledge the support from the United States Department of Energy, Advanced Research Projects Agency – Energy (ARPA-E) NEXTCAR project (Award Number DE-AR0000794).
	\end{ack}
	
	\bibliography{ifacconf}             % bib file to produce the bibliography

\begin{thebibliography}{27}
\providecommand{\natexlab}[1]{#1}
\providecommand{\url}[1]{\texttt{#1}}
\providecommand{\urlprefix}{URL }
\expandafter\ifx\csname urlstyle\endcsname\relax
  \providecommand{\doi}[1]{doi:\discretionary{}{}{}#1}\else
  \providecommand{\doi}{doi:\discretionary{}{}{}\begingroup
  \urlstyle{rm}\Url}\fi

\bibitem[{Alam and McNabola(2014)}]{alam2014critical}
Alam, M.S. and McNabola, A. (2014).
\newblock A critical review and assessment of eco-driving policy \& technology:
  Benefits \& limitations.
\newblock \emph{Transport Policy}, 35, 42--49.

\bibitem[{Altch{\'e} and de~La~Fortelle(2017)}]{altche2017lstm}
Altch{\'e}, F. and de~La~Fortelle, A. (2017).
\newblock An lstm network for highway trajectory prediction.
\newblock In \emph{2017 IEEE 20th International Conference on Intelligent
  Transportation Systems (ITSC)}, 353--359. IEEE.

\bibitem[{Brackstone and McDonald(1999)}]{brackstone1999car}
Brackstone, M. and McDonald, M. (1999).
\newblock Car-following: a historical review.
\newblock \emph{Transportation Research Part F: Traffic Psychology and
  Behaviour}, 2(4), 181--196.

\bibitem[{Cho et~al.(2014)Cho, Van~Merri{\"e}nboer, Gulcehre, Bahdanau,
  Bougares, Schwenk, and Bengio}]{cho2014learning}
Cho, K., Van~Merri{\"e}nboer, B., Gulcehre, C., Bahdanau, D., Bougares, F.,
  Schwenk, H., and Bengio, Y. (2014).
\newblock Learning phrase representations using rnn encoder-decoder for
  statistical machine translation.
\newblock \emph{arXiv preprint arXiv:1406.1078}.

\bibitem[{Chung et~al.(2014)Chung, Gulcehre, Cho, and
  Bengio}]{chung2014empirical}
Chung, J., Gulcehre, C., Cho, K., and Bengio, Y. (2014).
\newblock Empirical evaluation of gated recurrent neural networks on sequence
  modeling.
\newblock \emph{arXiv preprint arXiv:1412.3555}.

\bibitem[{Firoozi et~al.(2018)Firoozi, Nazari, Guanetti, O'Gorman, and
  Borrelli}]{firoozi2018safe}
Firoozi, R., Nazari, S., Guanetti, J., O'Gorman, R., and Borrelli, F. (2018).
\newblock Safe adaptive cruise control with road grade preview and v2v
  communication.
\newblock \emph{arXiv preprint arXiv:1810.09000}.

\bibitem[{Gao et~al.(2020)Gao, Huang, Zhang, Han, Wang, Zhang, and
  Lin}]{gao2020short}
Gao, S., Huang, Y., Zhang, S., Han, J., Wang, G., Zhang, M., and Lin, Q.
  (2020).
\newblock Short-term runoff prediction with gru and lstm networks without
  requiring time step optimization during sample generation.
\newblock \emph{Journal of Hydrology}, 589, 125188.

\bibitem[{Guanetti et~al.(2018)Guanetti, Kim, and
  Borrelli}]{guanetti2018control}
Guanetti, J., Kim, Y., and Borrelli, F. (2018).
\newblock Control of connected and automated vehicles: State of the art and
  future challenges.
\newblock \emph{Annual reviews in control}, 45, 18--40.

\bibitem[{Gupta(2019)}]{gupta2019thesis}
Gupta, S. (2019).
\newblock \emph{Look-Ahead Optimization of a Connected and Automated 48V
  Mild-Hybrid Electric Vehicle}.
\newblock Master thesis, The Ohio State University, USA.

\bibitem[{Gupta et~al.(2019)Gupta, Deshpande, Tulpule, Canova, and
  Rizzoni}]{gupta2019enhanced}
Gupta, S., Deshpande, S.R., Tulpule, P., Canova, M., and Rizzoni, G. (2019).
\newblock An enhanced driver model for evaluating fuel economy on real-world
  routes.
\newblock \emph{IFAC-PapersOnLine}, 52(5), 574--579.

\bibitem[{Gupta et~al.(2020)Gupta, Deshpande, Tufano, Canova, Rizzoni, Aggoune,
  Olin, and Kirwan}]{gupta2020estimation}
Gupta, S., Deshpande, S.R., Tufano, D., Canova, M., Rizzoni, G., Aggoune, K.,
  Olin, P., and Kirwan, J. (2020).
\newblock Estimation of fuel economy on real-world routes for next-generation
  connected and automated hybrid powertrains.
\newblock Technical report, SAE Technical Paper.

\bibitem[{Guzzella et~al.(2007)Guzzella, Sciarretta
  et~al.}]{guzzella2007vehicle}
Guzzella, L., Sciarretta, A., et~al. (2007).
\newblock \emph{Vehicle propulsion systems}, volume~1.
\newblock Springer.

\bibitem[{Hochreiter and Schmidhuber(1997)}]{hochreiter1997long}
Hochreiter, S. and Schmidhuber, J. (1997).
\newblock Long short-term memory.
\newblock \emph{Neural computation}, 9(8), 1735--1780.

\bibitem[{Hyeon et~al.(2019)Hyeon, Kim, Prakash, and
  Stefanopoulou}]{hyeon2019influence}
Hyeon, E., Kim, Y., Prakash, N., and Stefanopoulou, A.G. (2019).
\newblock Influence of speed forecasting on the performance of ecological
  adaptive cruise control.
\newblock In \emph{ASME 2019 Dynamic Systems and Control Conference}. American
  Society of Mechanical Engineers Digital Collection.

\bibitem[{Hyeon et~al.(2021)Hyeon, Shen, Karbowski, and
  Rousseau}]{hyeon2021forecasting}
Hyeon, E., Shen, D., Karbowski, D., and Rousseau, A. (2021).
\newblock Forecasting short to mid-length speed trajectories of preceding
  vehicle using v2x connectivity for eco-driving of electric vehicles.
\newblock Technical report, SAE Technical Paper.

\bibitem[{Kamal et~al.(2015)Kamal, Taguchi, and
  Yoshimura}]{kamal2015intersection}
Kamal, M.A.S., Taguchi, S., and Yoshimura, T. (2015).
\newblock Intersection vehicle cooperative eco-driving in the context of
  partially connected vehicle environment.
\newblock In \emph{2015 IEEE 18th International Conference on Intelligent
  Transportation Systems}, 1261--1266. IEEE.

\bibitem[{Kovvali et~al.(2007)Kovvali, Alexiadis, and
  Zhang~PE}]{kovvali2007video}
Kovvali, V.G., Alexiadis, V., and Zhang~PE, L. (2007).
\newblock Video-based vehicle trajectory data collection.
\newblock Technical report.

\bibitem[{Lef{\`e}vre et~al.(2014)Lef{\`e}vre, Sun, Bajcsy, and
  Laugier}]{lefevre2014comparison}
Lef{\`e}vre, S., Sun, C., Bajcsy, R., and Laugier, C. (2014).
\newblock Comparison of parametric and non-parametric approaches for vehicle
  speed prediction.
\newblock In \emph{2014 American Control Conference}, 3494--3499. IEEE.

\bibitem[{Olin et~al.(2019)Olin, Aggoune, Tang, Confer, Kirwan, Deshpande,
  Gupta, Tulpule, Canova, and Rizzoni}]{olin2019reducing}
Olin, P., Aggoune, K., Tang, L., Confer, K., Kirwan, J., Deshpande, S.R.,
  Gupta, S., Tulpule, P., Canova, M., and Rizzoni, G. (2019).
\newblock Reducing fuel consumption by using information from connected and
  automated vehicle modules to optimize propulsion system control.
\newblock Technical report, SAE Technical Paper.

\bibitem[{Pascanu et~al.(2013)Pascanu, Mikolov, and
  Bengio}]{pascanu2013difficulty}
Pascanu, R., Mikolov, T., and Bengio, Y. (2013).
\newblock On the difficulty of training recurrent neural networks.
\newblock In \emph{International conference on machine learning}, 1310--1318.
  PMLR.

\bibitem[{Rajakumar~Deshpande et~al.(2020)Rajakumar~Deshpande, Gupta, Kibalama,
  Pivaro, and Canova}]{rajakumar2020benchmarking}
Rajakumar~Deshpande, S., Gupta, S., Kibalama, D., Pivaro, N., and Canova, M.
  (2020).
\newblock Benchmarking fuel economy of connected and automated vehicles in real
  world driving conditions via monte carlo simulation.
\newblock In \emph{Dynamic Systems and Control Conference}, volume 84270,
  V001T10A004. American Society of Mechanical Engineers.

\bibitem[{Sun et~al.(2014)Sun, Hu, Moura, and Sun}]{sun2014velocity}
Sun, C., Hu, X., Moura, S.J., and Sun, F. (2014).
\newblock Velocity predictors for predictive energy management in hybrid
  electric vehicles.
\newblock \emph{IEEE Transactions on Control Systems Technology}, 23(3),
  1197--1204.

\bibitem[{Sutskever et~al.(2014)Sutskever, Vinyals, and
  Le}]{sutskever2014sequence}
Sutskever, I., Vinyals, O., and Le, Q.V. (2014).
\newblock Sequence to sequence learning with neural networks.
\newblock \emph{arXiv preprint arXiv:1409.3215}.

\bibitem[{Toledo et~al.(2007)Toledo, Koutsopoulos, and
  Ben-Akiva}]{toledo2007integrated}
Toledo, T., Koutsopoulos, H.N., and Ben-Akiva, M. (2007).
\newblock Integrated driving behavior modeling.
\newblock \emph{Transportation Research Part C: Emerging Technologies}, 15(2),
  96--112.

\bibitem[{Xu and Peng(2018)}]{xu2018design}
Xu, S. and Peng, H. (2018).
\newblock Design and comparison of fuel-saving speed planning algorithms for
  automated vehicles.
\newblock \emph{IEEE Access}, 6, 9070--9080.

\bibitem[{Zhang et~al.(2017)Zhang, Du, and Dai}]{zhang2017gru}
Zhang, J., Du, J., and Dai, L. (2017).
\newblock A gru-based encoder-decoder approach with attention for online
  handwritten mathematical expression recognition.
\newblock In \emph{2017 14th IAPR International Conference on Document Analysis
  and Recognition (ICDAR)}, volume~1, 902--907. IEEE.

\bibitem[{Zhu et~al.(2021)Zhu, Gupta, Pivaro, Deshpande, and
  Canova}]{zhu2021gpu}
Zhu, Z., Gupta, S., Pivaro, N., Deshpande, S.R., and Canova, M. (2021).
\newblock A gpu implementation of a look-ahead optimal controller for
  eco-driving based on dynamic programming.
\newblock \emph{arXiv preprint arXiv:2104.01284}.

\end{thebibliography}

\end{document}